\documentclass[aps,prb,superscriptaddress,twocolumn]{revtex4-1}

\usepackage{graphics}
\usepackage{graphicx} 
\usepackage{dcolumn}
\usepackage{bm}

\begin{document}


\title{Evolution of competing magnetic order in the  ${J}_{\mathrm{eff}}\mathbf{=}1/2$  insulating state of Sr$_2$Ir$_{\rm 1-x}$Ru$\rm _x$O$_4$}



\author{S.~Calder}
\email{caldersa@ornl.gov}
\affiliation{Quantum Condensed Matter Division, Oak Ridge National Laboratory, Oak Ridge, TN 37831.}

\author{J. W.~Kim}
\affiliation{Advanced Photon Source, Argonne National Laboratory, Argonne, IL 60439.}

\author{G.-X.~Cao}
\affiliation{Department of Materials Science and Engineering, University of Tennessee, Knoxville, TN 37996.}
\affiliation{Materials Science and Technology Division, Oak Ridge National Laboratory, Oak Ridge, TN 37831.}

\author{C.~Cantoni}
\affiliation{Materials Science and Technology Division, Oak Ridge National Laboratory, Oak Ridge, TN 37831.}

\author{A.~F.~May}
\affiliation{Materials Science and Technology Division, Oak Ridge National Laboratory, Oak Ridge, TN 37831.}

\author{H.~B.~Cao}
\affiliation{Quantum Condensed Matter Division, Oak Ridge National Laboratory, Oak Ridge, TN 37831.}

\author{A.~A.~Aczel}
\affiliation{Quantum Condensed Matter Division, Oak Ridge National Laboratory, Oak Ridge, TN 37831.}

\author{M.~Matsuda}
\affiliation{Quantum Condensed Matter Division, Oak Ridge National Laboratory, Oak Ridge, TN 37831.}

\author{Y.~Choi}
\affiliation{Advanced Photon Source, Argonne National Laboratory, Argonne, IL 60439.}

\author{D.~Haskel}
\affiliation{Advanced Photon Source, Argonne National Laboratory, Argonne, IL 60439.}

\author{B.~C.~Sales}
\affiliation{Materials Science and Technology Division, Oak Ridge National Laboratory, Oak Ridge, TN 37831.}

\author{D.~Mandrus}
\affiliation{Department of Materials Science and Engineering, University of Tennessee, Knoxville, TN 37996.}
\affiliation{Materials Science and Technology Division, Oak Ridge National Laboratory, Oak Ridge, TN 37831.}

\author{M.~D.~Lumsden}
\affiliation{Quantum Condensed Matter Division, Oak Ridge National Laboratory, Oak Ridge, TN 37831.}
 
\author{A.~D.~Christianson}
\affiliation{Quantum Condensed Matter Division, Oak Ridge National Laboratory, Oak Ridge, TN 37831.}

 

\begin{abstract}
We investigate the magnetic properties of the series Sr$_2$Ir$\rm _{1-x}$Ru$\rm _{x}$O$_4$ with neutron, resonant x-ray and magnetization measurements. The results indicate an evolution and coexistence of magnetic structures via a spin flop transition from ab-plane  to c-axis collinear order as the 5d Ir$^{4+}$ ions are replaced with an increasing concentration of 4d Ru$^{4+}$ ions. The magnetic structures within the ordered regime of the phase diagram (x$<$0.3) are reported. Despite the changes in magnetic structure no alteration of the ${J}_{\mathrm{eff}}\mathbf{=}1/2$ ground state is observed. The behavior of Sr$_2$Ir$\rm _{1-x}$Ru$\rm _{x}$O$_4$ is consistent with electronic phase separation and diverges from a standard scenario of hole doping. The role of lattice alterations with doping on the magnetic and insulating behavior is considered. The results presented here provide insight into the magnetic insulating states in strong spin-orbit coupled materials and the role perturbations play in altering the behavior.
 \end{abstract}

\maketitle

\section{\label{sec:Introduction}Introduction}

Investigations of materials with 5d transition metal ions have opened up new paradigms in condensed matter physics. In this regime spin-orbit coupling (SOC) can play a prominent role by competing directly with several phenomena, such as electron correlations, lattice alterations, bandwidth and crystal field splitting \cite{annurev-conmatphys-020911-125138}. The behavior manifested due to these finely balanced interactions is often novel and diverse and allows 5d systems to span a wide phase space containing metals and insulators with exotic behavior such as Weyl semimetals, axion insulators and novel metal-insulator transitions \cite{PhysRevB.83.205101, NaturePesin,annurev-conmatphys-020911-125138}. Moreover robust magnetism often emerges, despite the apparent obstacle of reduced correlation and itinerant nature of 5d ions compared to analogous 3d based systems. 

Particular focus in 5d systems with strong SOC has centered on iridates containing Ir$^{4+}$ ions \cite{annurev-conmatphys-020911-125138}. These systems can host unusual magnetic insulating states  where the SOC splits the non-degenerate 5d$^5$ t$_{2g}$  manifold into a fully occupied ${J}_{\mathrm{eff}}\mathbf{=}3/2$ manifold and a half-filled ${J}_{\mathrm{eff}}\mathbf{=}1/2$  shell that can be further split by the on-site Coulomb interaction. The result is a magnetic ${J}_{\mathrm{eff}}\mathbf{=}1/2$ SOC Mott-like insulating state. Experimental evidence was initially reported in Sr$_2$IrO$_4$ \cite{KimScience}, followed by an increasing number of Ir$^{4+}$ based transition metal oxides \cite{PhysRevLett.109.037204, PhysRevB.89.081104,PhysRevB.87.155136,PhysRevLett.110.117207,PhysRevLett.113.247601}. 

The role of magnetism and lattice effects, notably the existence of the ${J}_{\mathrm{eff}}\mathbf{=}1/2$ state despite significant non-cubic distortions, have continued to prompt debate as to the nature of the ground state and subsequent emergent properties. In this investigation we perturb Sr$_2$IrO$_4$ via  chemical substitution. Both electron doping and hole doping Sr$_2$IrO$_4$ are interesting avenues that have undergone limited investigations. Indeed superconductivity has been postulated to occur via electron doping on the Sr site\cite{PhysRevLett.106.136402} and experimental evidence of the proximity of the parent compound to a superconducting regime has been proposed due to analogous spin excitations with the parent cuprates \cite{PhysRevLett.108.177003}. Conversely hole or electron doping on the Ir site offers a handle to control the  onsite and intra-site interactions of the magnetic ion responsible for the ${J}_{\mathrm{eff}}\mathbf{=}1/2$ state. In this investigation we follow the latter route and investigate the series  Sr$_2$Ir$\rm _{1-x}$Ru$\rm _{x}$O$_4$ over the full magnetically ordered regime and into the unordered state. We verify this corresponds to hole doping with the substitution of 4d$^4$ Ru$^{4+}$ ions on the 5d$^5$ Ir$^{4+}$ site. A key focus of our investigation is the novel magnetism in the strong SOC limit. In this regime new physics can emerge due to SOC allowing the mixing of orbitals, where symmetry would usually prohibit such an occurrence. This can lead to the presence of dominant anisotropic rather than isotropic magnetic exchange couplings in the ground state in the form of Kitaev interactions \cite{PhysRevLett.102.017205}.

The end members of the Sr$_2$Ir$\rm _{1-x}$Ru$\rm _{x}$O$_4$ series exhibit distance physical behavior. Sr$_2$RuO$_4$ is a nearly ferromagnetic metal which exhibits unconventional p-wave superconductivity below 1.5 K attributed to p-wave pairing \cite{maenonature}. The electronic configuration of the Ru$^{4+}$ ion results in S=1 in contrast to the ${J}_{\mathrm{eff}}\mathbf{=}1/2$ antiferromagnetic insulating state in Sr$_2$IrO$_4$. Therefore Sr$_2$RuO$_4$ (metallic) and Sr$_2$IrO$_4$ (insulator) reside on opposite sides of a metal-insulator divide. Both Sr$_2$RuO$_4$ and Sr$_2$IrO$_4$ form the K$_2$NiF$_4$-type structure, with Sr$_2$RuO$_4$ adopting the I4/mmm space group and Sr$_2$IrO$_4$ the I4$_1$/acd, although a recent report suggested the I4$_1$/a space group \cite{PhysRevLett.114.096404}. The difference between the space groups for Sr$_2$IrO$_4$ and Sr$_2$RuO$_4$  is the result of rotation of  of the IrO$_6$ octahedra. This series was previously investigated in Ref.~\onlinecite{PhysRevB.49.11890}, where x-ray diffraction  indicated the structural change for the series Sr$_2$Ir$\rm _{1-x}$Ru$\rm _{x}$O$_4$ to occur around x=0.7. That investigation was restricted to  powder samples and involved no microscopic probes of magnetism. Recently a Raman investigation measured single crystals of Sr$_2$Ir$\rm _{1-x}$Ru$\rm _{x}$O$_4$, with the measurements focusing on the octahedral rotations \cite{PhysRevB.89.104406}. Here we investigate single crystals with both neutron scattering and resonant magnetic x-ray scattering (RMXS) that allows us to probe the long range magnetic structure and nature of the electronic ground state of the Ir ion as a function of Ru doping and consider the role of the competing interactions in the system.

\section{\label{sec:ExptMethods}Experimental Methods}

Single crystals of Sr$_2$Ir$\rm _{1-x}$Ru$\rm _{x}$O$_4$ were grown in a Pt crucible using the flux method. Crystals of mm dimensions with masses ranging from 5 to 20 mg were produced for Ru concentrations up to $40 \%$.  Additionally powder samples of x=0.05 and 0.2 were prepared by standard solid state techniques. Neutron scattering was performed on the single crystals at the High Flux Isotope Reactor (HFIR) on HB-1, HB-1A and HB-3A. The triple axis instrument HB-1A was used in elastic mode with a wavelength of 2.36 $\rm \AA$ and collimation 40'-40'-open-open to determine the magnetic structure. Polarized neutron scattering measurements were performed on HB-1. Heusler monochromator and analyzer crystals were used to perform the polarized measurements with a guide field giving the option of flipping the spin in horizontal and vertical fields at the sample position. The beam was collimated with 48'-open-80'-open solar collimators. The crystal structure was measured on the Four-circle single crystal neutron diffractometer HB-3A with a wavelength of 1.003 $\rm \AA$. Measurements on the powder samples were carried out on the HB-2A powder diffractometer at HFIR using a wavelength of 1.54 $\rm \AA$. Resonant magnetic x-ray scattering (RMXS) measurements were performed on beamline 6-ID-B at the Advanced Photon Source (APS) on single crystals.  Measurements  were carried out at both the L$_2$ (12.824 keV) and L$_3$ (11.215 keV) resonant edges of iridium. Graphite was used as the polarization analyzer at the (0 0 10) and (0 0 8) reflections on the L$_2$ and L$_3$ edges, respectively, to achieve a scattering angle close to 90$^{\circ}$. An analysis of the photon polarization allowed magnetic and charge scattering to be distinguished. To observe the sample fluorescence, energy scans were performed without the analyzer and with the detector away from any Bragg peaks through both absorption energies. X-ray magnetic circular dichroism (XMCD) and X-ray absorption near edge spectroscopy (XANES) was performed at the Ir L-edge on beamline 4-ID-D at the APS. The XMCD measurements were performed in an $\pm$3 T field at 1.8 K. Powder samples were used to ensure uniform sample thickness and all measurements were performed in transmission mode. The sample magnetization M(T,H) was measured with a Quantum Design (QD) magnetic property measurement system (MPMS) with an applied field of 100 Oe. 

\section{\label{sec:ExptRresults}Results and analysis}

\subsection{\label{sec:XANES} Valence determination of Ir in Sr$_2$Ir$_{\rm 1-x}$Ru$\rm _x$O$_4$}

\begin{figure}[tb]
	\centering     
	\includegraphics[trim=2.2cm  1cm 3cm 5cm,clip=true, width=0.9\columnwidth]{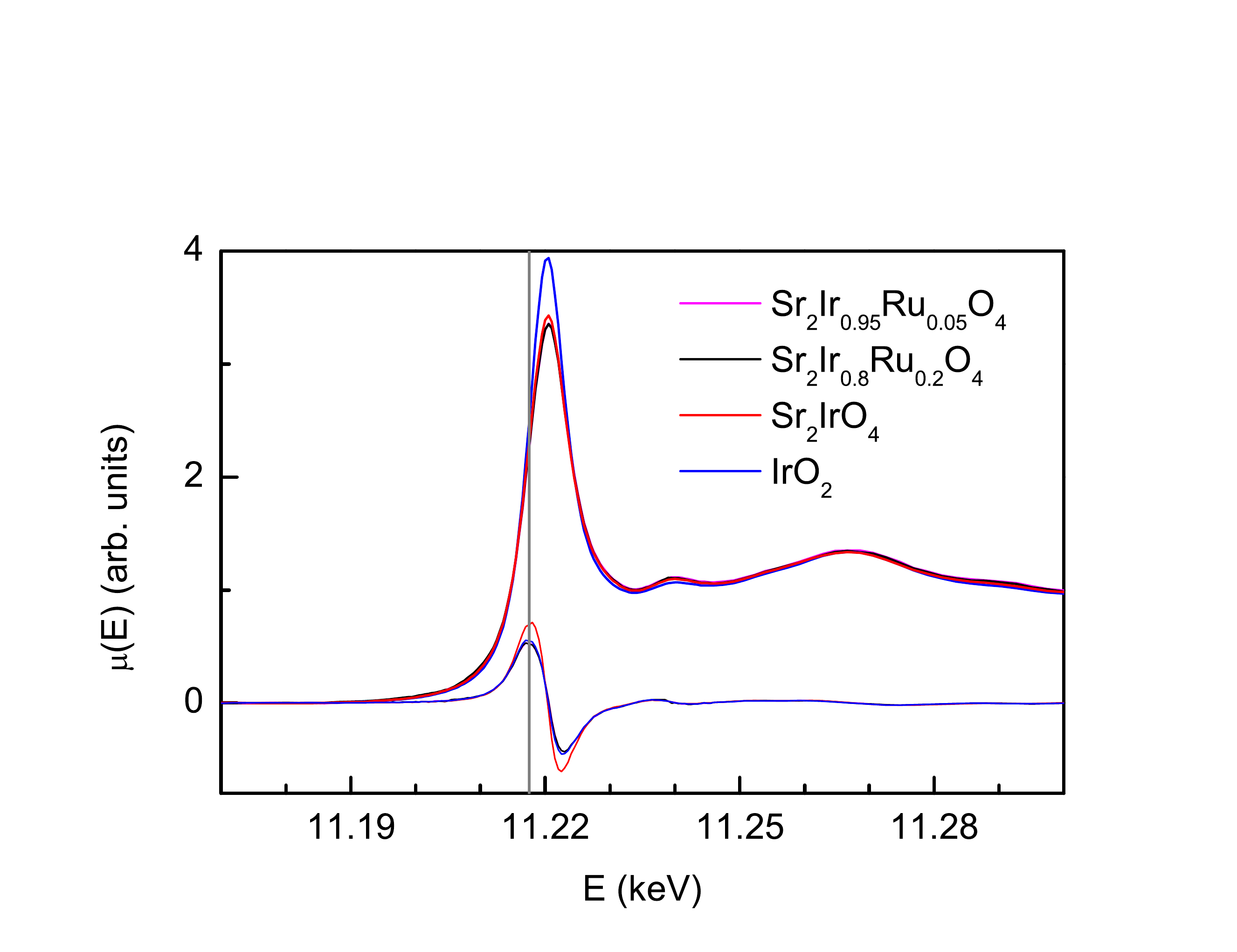} 
	\caption{\label{FigXANES} XANES measurements at the Ir L$_3$-edge for Sr$_2$Ir$\rm _{1-x}$Ru$\rm _{x}$O$_4$, with x=0, 0.05 and 0.2, and a IrO$_2$ standard. The white line absorption intensity and the derivative are both shown for each sample. The same absorption energy  of 11.2175 keV is observed for all the samples measured indicating a valence of Ir$^{4+}$.}
\end{figure}

We initially consider whether the introduction of Ru onto the Ir site alters the valence state of iridium. It is reasonable to assume that Ru adopts the Ru$^{4+}$ valence resulting in Sr$_2$Ir$^{4+}$$_{1-x}$Ru$^{4+}$$_x$O$_4$ for all values of x. However, a  similar reasoning proved incorrect in the series Sr$_2$Ir$_{1-x}$Rh$_x$O$_4$. There Rh formed Rh$^{3+}$ for x$>$0 resulting in mixed magnetic Ir$^{4+}$ and non-magnetic Ir$^{5+}$. \cite{PhysRevB.89.054409} To probe the Ir valence in Sr$_2$Ir$\rm _{1-x}$Ru$\rm _{x}$O$_4$ we performed XANES measurements on the x=0, 0.05 and 0.2 members of the series and compared this to an established iridium standard IrO$_2$. The XANES results are shown in Fig.~\ref{FigXANES} and show no indication of an altered Ir valance from 4+, unlike similar measurements for the Rh case where a pronounced shift in the energy of the resonant edge was observed \cite{PhysRevB.89.054409}. Results at the L$_2$ edge show a similar overlap in the energy position  of the white line as the L$_3$ edge. Therefore the series Sr$_2$Ir$^{4+}$$_{1-x}$Ru$^{4+}$$_x$O$_4$ corresponds to hole doping on the Ir site.

\subsection{\label{sec:bulk measurements} Magnetization measurements of Sr$_2$Ir$_{\rm 1-x}$Ru$\rm _x$O$_4$}
 
\begin{figure}[tb]
	\centering     
\end{figure}
 
We begin our magnetic investigation with magnetization measurements on single crystal samples  of Sr$_2$Ir$\rm _{1-x}$Ru$\rm _{x}$O$_4$ with x=0.05, 0.1, 0.2 and 0.3, shown in Fig.~\ref{FigMagnetization}. The results indicate a rather complicated magnetic temperature dependence with the x=0.05 and x=0.1 concentrations showing more than one anomaly. For example for x=0.05 there are two pronounced anomalies, one around 200 K and another at 150 K. Similarly x=0.1 has two anomalies, located at 125 K and 160 K. The magnetization measurements in Fig.~\ref{FigMagnetization} indicate long ranged magnetic order up to x=0.2. At least qualitatively the results are similar to those on powder samples presented in Ref.~\onlinecite{PhysRevB.49.11890} and therefore allows comparisons between our current investigation and the results previously obtained.

\subsection{\label{sec:magstruct} Magnetic structure of the series Sr$_2$Ir$_{\rm 1-x}$Ru$\rm _x$O$_4$}

\begin{figure}[tb]
   \centering                   
  \includegraphics[trim=0.25cm 0cm 0cm 0cm,clip=true, width=1.0\columnwidth]{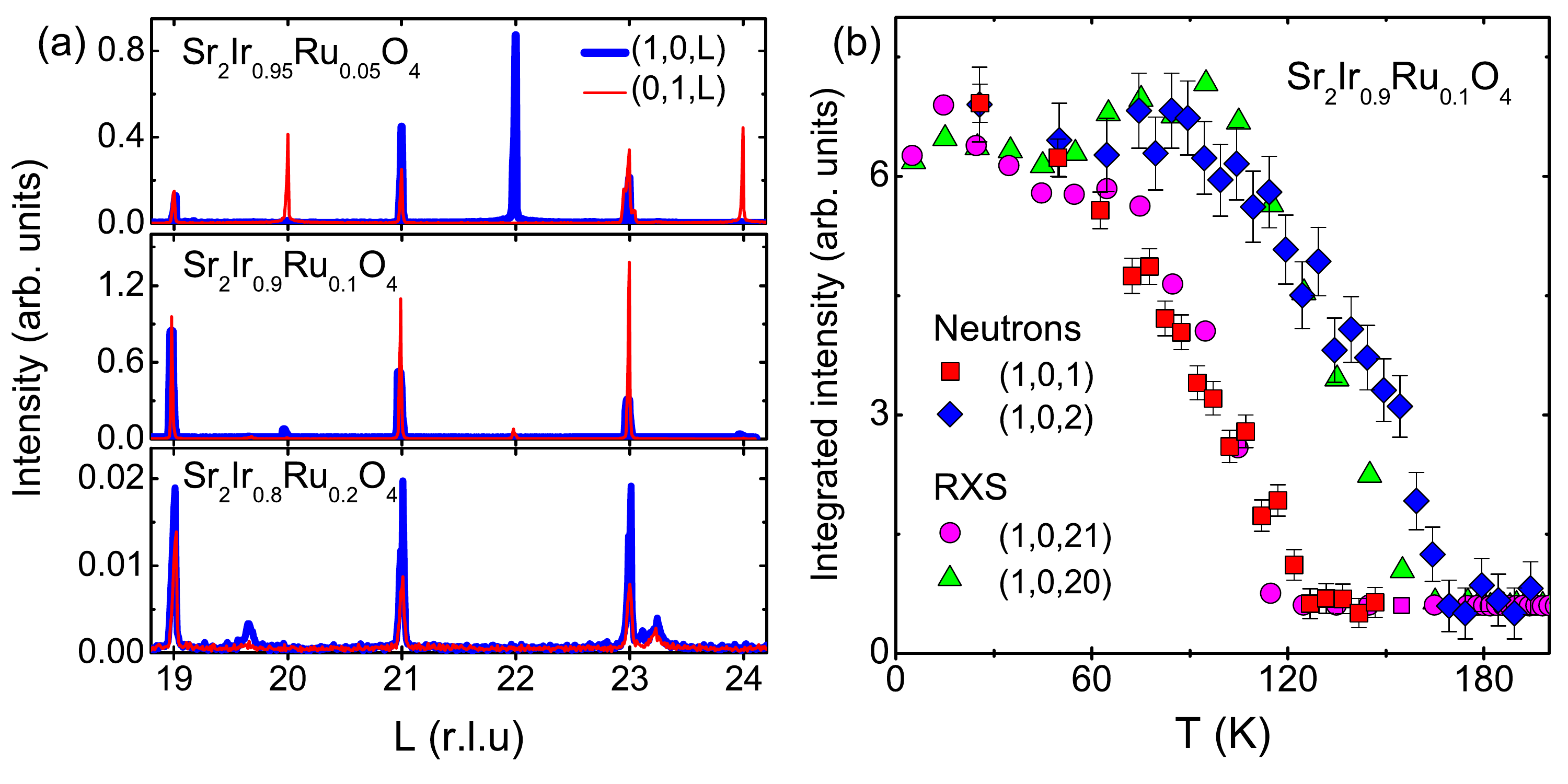}       
  \includegraphics[trim=0cm 0cm 0cm 0cm,clip=true, width=1.0\columnwidth]{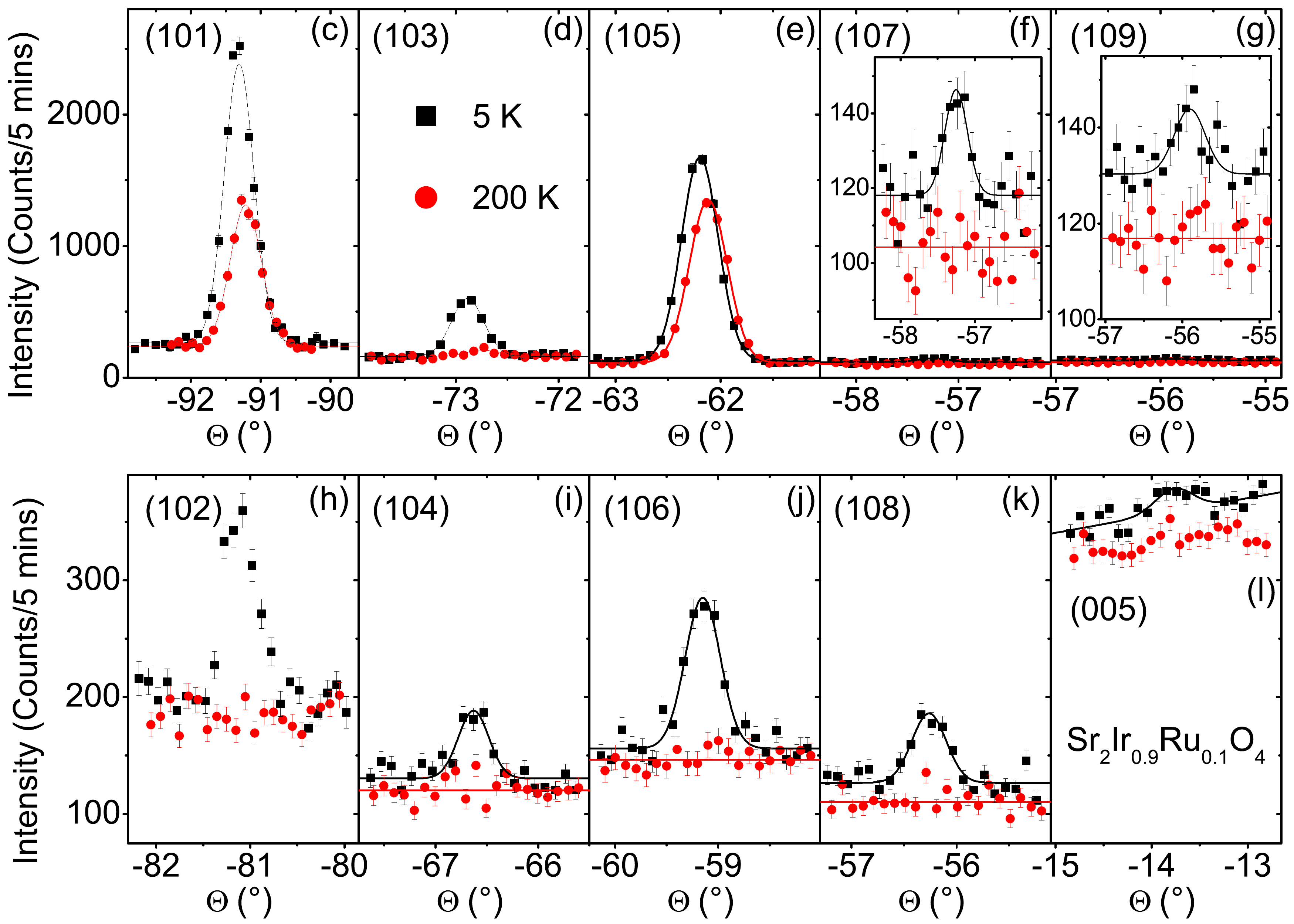}       
 \caption{\label{Figneutron_Ru0p1} (a) RMXS measurements on single crystals of  Sr$_2$Ir$\rm _{1-x}$Ru$\rm _{x}$O$_4$ with x=0.05, 0.1 and 0.2. (b) Comparison of the intensity of (1,0,$2n$+1) and (1,0,$2n$) magnetic reflections in Sr$_2$Ir$_{0.9}$Ru$_{0.1}$O$_4$ measured with both neutrons and RMXS. The neutron and x-ray measurements involved different crystals from the same batch. The intensities have been normalized to their respective backgrounds intensities and scaled so all reflections are in a single plot. (c)-(l) Neutron scattering measurements on a single crystal of  Sr$_2$Ir$_{0.9}$Ru$_{0.1}$O$_4$ for several reflections. Scattering is observed at (1,0,L) for both L=odd and L=even.}
\end{figure}

To probe the magnetic structure we combine neutron and RMXS measurements. Results for RMXS measurements at the Iridium L$_3$-edge are shown in Fig.~\ref{Figneutron_Ru0p1}(a)-(b) for Sr$_2$Ir$\rm _{1-x}$Ru$\rm _{x}$O$_4$ single crystals with concentrations of x=0.05, 0.1 and 0.2. Magnetic scattering is observed at both (1,0,$2n$) and (1,0,$2n$+1) reflections for x=0.05. Substituting in x=0.1 produces the same magnetic reflections  at (1,0,$2n$) and (1,0,$2n$+1), however the intensity of the L=even reflections are much reduced compared to the L=odd reflections. Moving to x=0.2 only the (1,0,$2n$+1) magnetic reflections are present. 

This is distinctly different behavior from the few previous cases where the magnetic structure has been probed in doped Sr$_2$IrO$_4$. Undoped Sr$_2$IrO$_4$ has magnetic (1,0,$2n$) reflections with spins in the ab-plane \cite{KimScience,PhysRevB.87.140406,PhysRevB.87.144405}. For Mn-doping \cite{PhysRevB.86.220403}, Rh-doping \cite{PhysRevB.89.054409} or the application of an applied field \cite{KimScience}  to Sr$_2$IrO$_4$ only (1,0,$2n$+1) reflections are present, indicating an altered magnetic structure to the undoped case. 

\subsubsection{\label{sec:magstruct_Ru0p1} Magnetic structure of Sr$_2$Ir$_{0.9}$Ru$_{0.1}$O$_4$}

Focusing on the Sr$_2$Ir$\rm _{1-x}$Ru$\rm _{x}$O$_4$ composition with x=0.1 that has both (1,0,$2n$) and (1,0,$2n$+1) reflections  we measured several Bragg peaks with neutrons and RMXS, see Fig.~\ref{Figneutron_Ru0p1}. Following specific reflections we find the magnetic order parameters reveal two distinct magnetic ordering temperatures, see Fig.~\ref{Figneutron_Ru0p1}(b). This is confirmed using both x-rays and neutrons on different samples. The (1,0,$2n$+1) reflections develop at 120 K, whereas (1,0,$2n$) reflections appear at160 K. The ordering of the (1,0,$2n$+1) reflections appear to not be reflected by any associated anomaly in the (1,0,$2n$) order parameter, suggesting a possible decoupling of two magnetic phases within the sample.  Susceptibility measurements confirm a change in magnetization at these temperatures for Sr$_2$Ir$_{0.9}$Ru$_{0.1}$O$_4$ on a crystal from the same batch (Fig.~\ref{FigMagnetization}), supporting the magnetic origin.

\begin{figure}[tb]
   \centering                   
   \includegraphics[trim=0cm 0cm 0cm 0cm,clip=true, width=1.0\columnwidth]{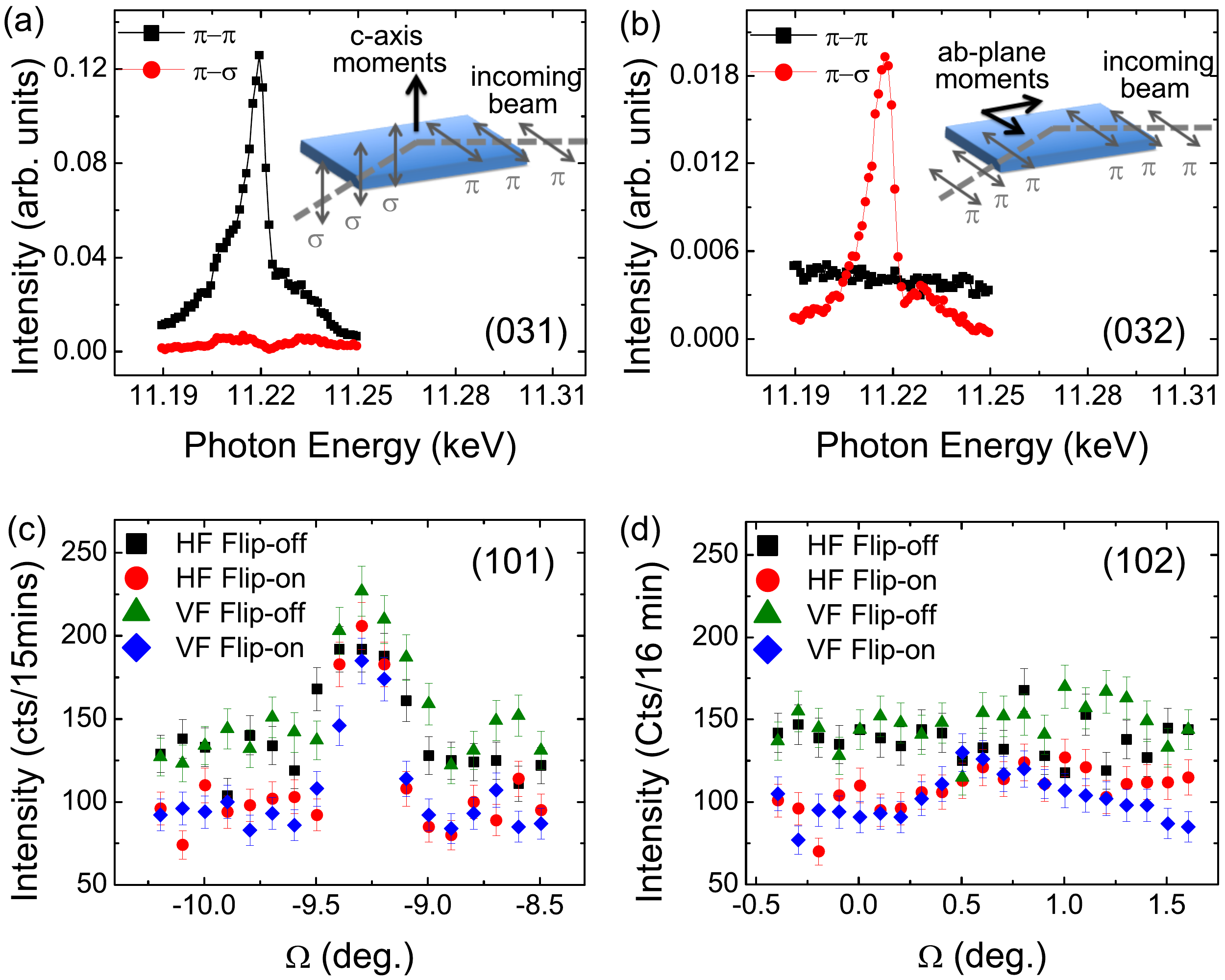} 
                   \caption{\label{Pol} By performing polarization analysis of (a)-(b) x-rays and (c)-(d) neutrons on Sr$_2$Ir$_{0.9}$Ru$_{0.1}$O$_4$ the spin direction associated with the magnetic reflections is determined. Both measurements were performed at 5 K.}
\end{figure}

In order to gain further understanding of the long range magnetic order in Sr$_2$Ir$_{0.9}$Ru$_{0.1}$O$_4$  we consider the polarization dependence of the scattering cross-section, firstly with x-rays and then with neutrons. A standard configuration to measure RMXS  in is a vertical geometry, however by measuring in a horizontal geometry we can utilize the polarization dependence of the incoming beam and analyzed beam to determine the spin direction of the ordered moments within the sample, see Fig.~\ref{Pol}(a)-(b). With the incoming beam $\pi$ polarized we find intensity for $\pi$-$\pi$ polarized analysis at the (0,3,1), equivalent to (1,0,$2n$+1) reflection, and no intensity for $\pi$-$\sigma$ polarization. This is consistent with spins oriented along the c-axis in the sample. Conversely for the (1,0,$2n$) reflection, measured at (0,3,2), we observe the opposite behavior with intensity only at $\pi$-$\sigma$ polarization that indicates the spins contributing to this reflection are confined to the ab-plane. This behavior was confirmed with polarized neutron scattering measurements on a single crystal sample of Sr$_2$Ir$_{0.9}$Ru$_{0.1}$O$_4$ at the (101) and (102) reflections. For the (101) reflections intensity is observed in all the four channels measured from the combinations of horizontal field (HF), vertical field (VF) and flip-on and flip-off, see Fig.~\ref{Pol}(c). This is consistent with both magnetic and nuclear contributions to the scattering, as expected given the observation of nuclear scattering at the (1 0 odd) reflections in the parent compound. Moreover with the crystal aligned in the (H0L) plane it indicates c-axis oriented spins since the purely magnetic intensity (HF flip-on) is similar to the magnetic intensity from the magnetic component perpendicular to the [101] direction in the (H0L) scattering plane (VF flip-on), which is almost along the c axis. Conversely  the behavior of the (102) reflection is distinctly different, as shown in Fig..~\ref{Pol}(d). The purely magnetic intensity (HF flip-on), though reduced, is roughly similar to the magnetic intensity mostly from the magnetic component along the b axis (VF, flip-off), in-line with the conclusions from the resonant x-ray results. 

Following a representational analysis approach, there is no  single magnetic structure consistent with a combination of both (1,0,$2n$+1) and (1,0,$2n$) reflections. We therefore considered the possibility of structural phase separation as an explanation for the apparent result of two different sets of magnetic reflections. We carefully measured the two theta dependence with both neutrons and x-rays of several reflections and within the resolution found only one reflection. This is consistent with a single structural domain in the single crystals samples studied here. This is based on the assumption that the two structural phases are epitaxial and would be accessible in the specific orientation of the aligned crystal. While this is a reasonable assumption  given that the magnetic reflections are commensurate with the lattice we considered the phase separation scenario with further dedicated measurements.

A desktop scanning electron microscope (SEM) with energy dispersive spectroscopy (EDS) was used to verify no macroscopic phase separation was present in the samples. Scanning transmission electron microscope (STEM) was then used to examine the local structure. The single crystals of both doped and undoped Sr$_2$IrO$_4$ were examined in plane view geometry using high-resolution Z-contrast scanning transmission electron microscopy (Z-STEM). The samples were found to have no structural defects and consequently our chemical and structural analysis does not reveal structural phase separation associated with Ru doping.

\begin{figure}[tb]
   \centering                   
            \includegraphics[trim=1.0cm 0.5cm 0.6cm 2cm,clip=true, width=0.69\columnwidth]{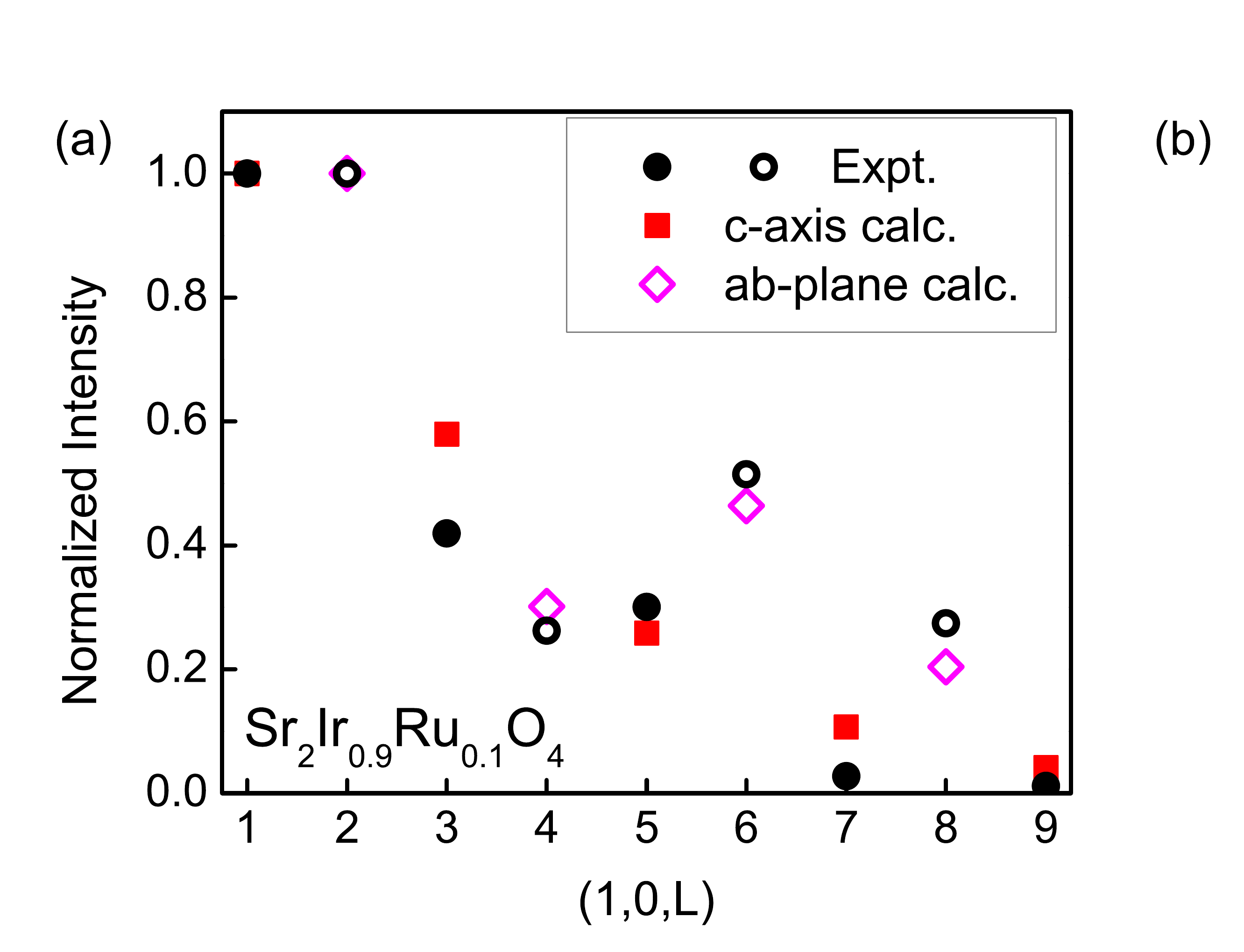}        
    \includegraphics[trim=7.5cm 7cm 7.5cm 7cm,clip=true, width=0.27\columnwidth]{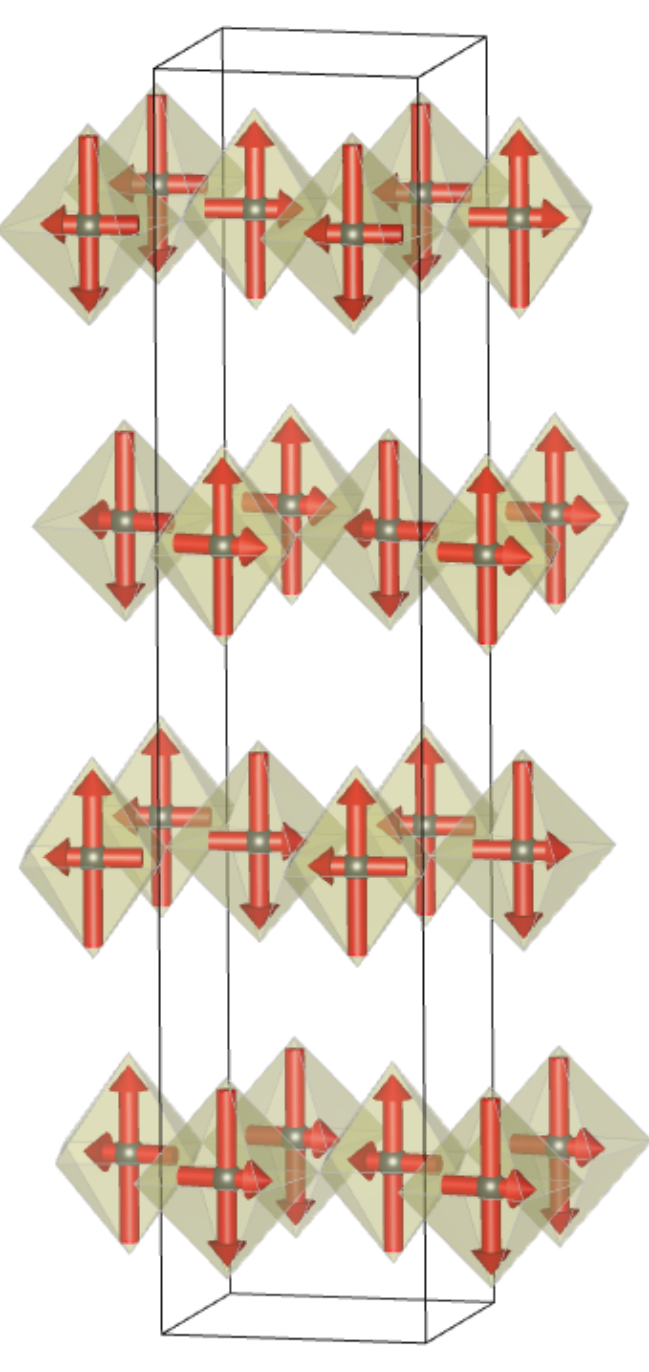} 
                   \caption{\label{Mag_Str_Ru0p1} (a) Experimental neutron scattering intensities for the magnetic reflections in Sr$_2$Ir$_{0.9}$Ru$_{0.1}$O$_4$ are compared with calculated magnetic structures for the case of spins in the ab-plane and c-axis. (b) Corresponding magnetic structures within the nuclear unit cell.}
\end{figure}

Using a representational analysis approach to find the long ranged ordered magnetic structure from our neutron results for Sr$_2$Ir$_{0.9}$Ru$_{0.1}$O$_4$ (Fig.~\ref{Figneutron_Ru0p1}(c)-(l)),  we  analyze the L=odd and L=even reflections separately. The experimental (1 0 even) reflections intensity  for L=2,4,6,8 is shown in Fig.~\ref{Mag_Str_Ru0p1}(a), normalized to the (1,0,2) intensity. The results are qualitatively the same as the Sr$_2$IrO$_4$ undoped case presented by both Refs.~\onlinecite{PhysRevB.87.140406} and \onlinecite{PhysRevB.87.144405}. To confirm this we modeled the magnetic structure, taking into account the lorentzian function correction using ResLib \cite{ResLib}, absorption of the sample and magnetic form factor for Ir$^{4+}$ using the information in Ref.~\onlinecite{Kobayashi:kx5002}. The experimental and calculated results are shown in Fig.~\ref{Mag_Str_Ru0p1}(a), along with the corresponding magnetic spin structure in Fig.~\ref{Mag_Str_Ru0p1}(b). Close agreement is found between the experimental and calculated magnetic intensities indicating that this indeed corresponds to the magnetic ordering that results in the (1 0 even) reflections. In terms of representational analysis this corresponds to $\Gamma_1$, with a propagation vector k=(111), for the Ir ion at the (0.5,0.25,0.125) position. Turning now to the (1 0 odd) reflections and following the polarization dependence that indicates c-axis aligned spins we model the magnetic structure. The experimental and calculated results are shown in  Fig.~\ref{Mag_Str_Ru0p1}(a) for L=1,3,5,7,9. This corresponds to the $\Gamma_1$ irreducible representation, with a propagation vector k=(000), for the Ir ion at the (0.5,0.25,0.125) position, as found for the case of Mn-doped Sr$_2$IrO$_4$ \cite{PhysRevB.86.220403}. Again close agreement is found between the intensity of the experimental and calculated magnetic reflections.    


Both magnetic structures, see Fig.~\ref{Mag_Str_Ru0p1}(b), are related by a spin flop from the ab-plane to the c-axis. Given the apparent occurrence of these two competing magnetic structures and no evidence for structural phase separation we assign the behavior as being due to electronic phase separation. This conclusion is in line with that arrived at from the Raman investigation of  Sr$_2$Ir$_{1-x}$Ru$_{x}$O$_4$ \cite{PhysRevB.89.104406} and also the Ru doping behavior in the related iridate Sr$_3$Ir$_2$O$_7$ \cite{DhittalSr327Ru}. We note, however, the intriguing similarity of the magnetic structure to the ortho-G-AF phase presented in Ref.~\onlinecite{PhysRevB.87.064407} that emerges from the Kugel-Khomskii model. For such a magnetic phase to exist would require weak nearest neighbor interactions and appreciable second and third neighbor interactions and it remains unclear if such a phase would exist in this SOC dominated system \cite{PhysRevLett.102.017205}.

\subsubsection{\label{sec:magstruct_Ru0p2} Magnetic structure of Sr$_2$Ir$_{0.8}$Ru$_{0.2}$O$_4$}

\begin{figure}[tb]
   \centering      
            \includegraphics[trim=1.8cm 3.5cm 4.5cm 0.2cm,clip=true, width=0.52\columnwidth]{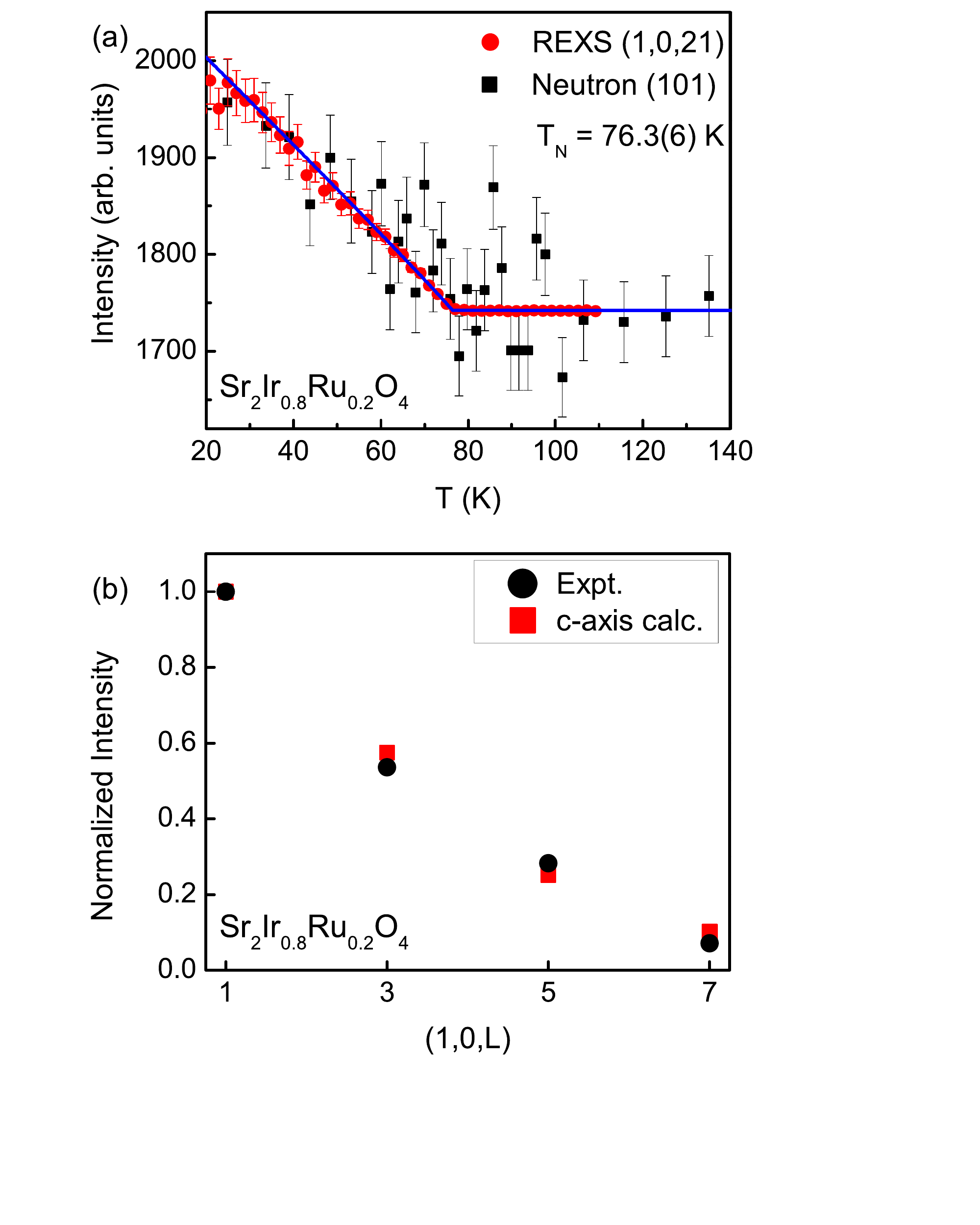}                    
              \includegraphics[trim=6cm 6.0cm 6.8cm 5.8cm,clip=true, width=0.42\columnwidth]{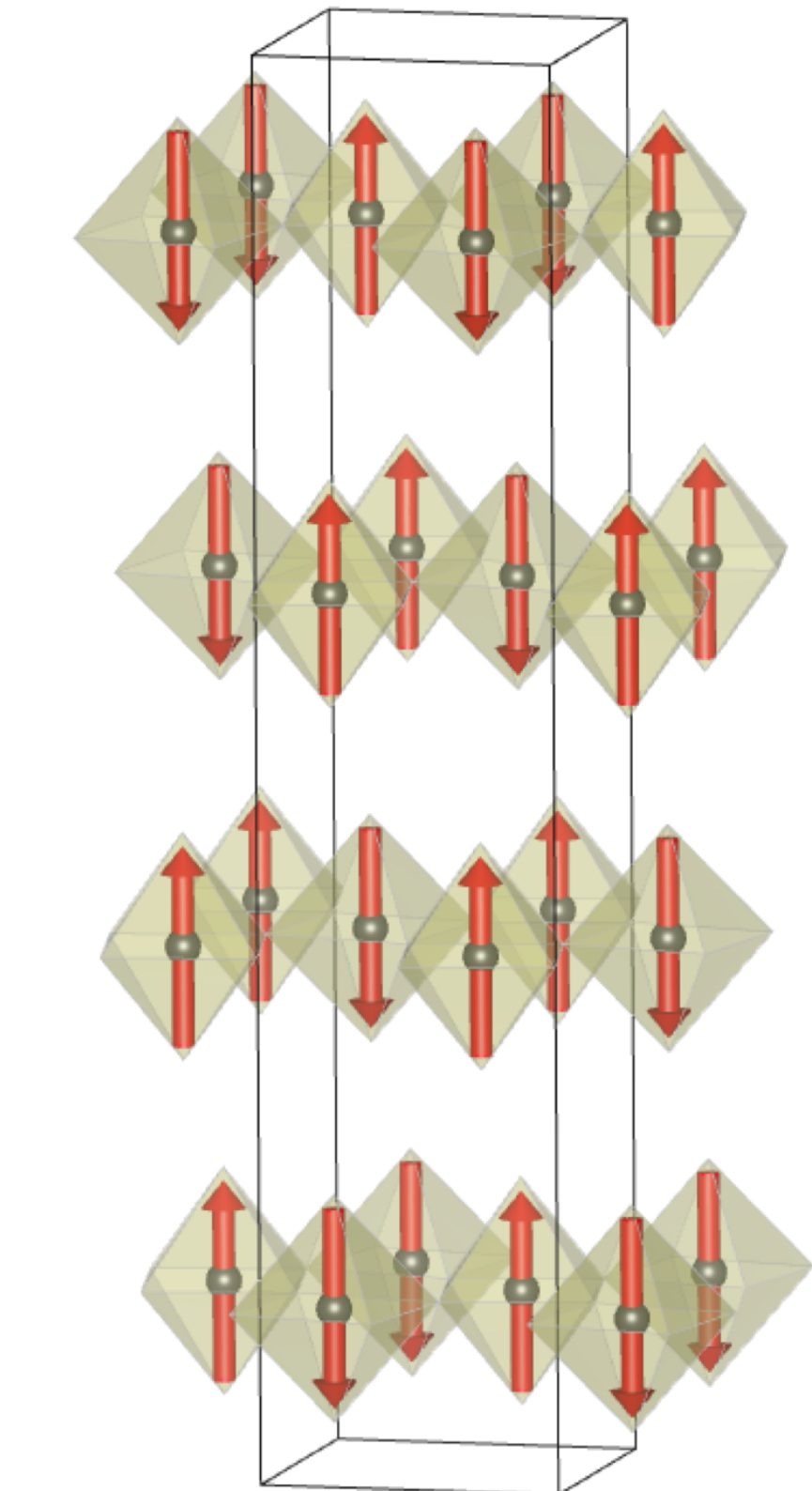} 
                   \caption{\label{Mag_Str_Ru0p2} (a) Intensity dependence of the (1 0 odd) magnetic reflection measured with neutrons at (101) and RMXS at (1,0,21). The RXMS results are fit to a power law and give a transition temperature at T$\rm _N$=76.3(6). (b) Experimental intensities for magnetic reflections compared to the calculated intensities for c-axis aligned spins. (c) The magnetic structure for Sr$_2$Ir$_{0.8}$Ru$_{0.2}$O$_4$ with spins aligned along the c-axis.}
\end{figure}

Substituting in more Ru leads to the occurrence of only one magnetic transition for Sr$_2$Ir$_{0.8}$Ru$_{0.2}$O$_4$ at T$\rm _N$=76.3(6) K, see Fig.~\ref{Mag_Str_Ru0p2}(a). We measured the (101),(103),(105) and (107) reflections and obtained the intensities shown in Fig.~\ref{Mag_Str_Ru0p2}(b). We again note that an additional weak nuclear contribution is present at (1 0 odd) reflections. Following the same method as for the (1 0 odd) case for the 10$\%$ Ru substitution we modeled the magnetic structure with spins along the c-axis, this structure is shown in Fig.~\ref{Mag_Str_Ru0p2}(c). The experimental and calculated intensities are compared in Fig.~\ref{Mag_Str_Ru0p2}(b) and show close agreement.  The ordered magnetic moment on the Ir ion is determined by scaling the magnetic intensities with measured nuclear reflections intensities. The ordered magnetic moment is 0.13(2)$\rm \mu_B$/Ir. Going to higher concentrations of Sr$_2$Ir$_{0.7}$Ru$_{0.3}$O$_4$  no long ranged magnetic order is observed.

\subsection{\label{sec:InsulatingState} Insulating state in Sr$_2$Ir$_{\rm 1-x}$Ru$\rm _x$O$_4$}


We now turn to consider the nature of the insulating state within the magnetically ordered regime. One of the principle experimental proofs of the SOC driven  ${J}_{\mathrm{eff}}\mathbf{=}1/2$ Mott insulating state in Sr$_2$IrO$_4$ was proposed on the basis of RMXS measurements \cite{KimScience}. It was argued that the observation of the large intensity at the L$_3$ and vanishing intensity at the L$_2$ edge was due to the alteration of the electronic ground state from a S=1/2 to a ${J}_{\mathrm{eff}}\mathbf{=}1/2$ scenario on the basis of the different 2p-5d transitions that are involved in the two different L-edges probed with RMXS. Subsequently several investigations have proceeded along the same route and used the L2:L3 branching ratio as evidence for a ${J}_{\mathrm{eff}}\mathbf{=}1/2$ state. However, it has been argued that for the case of spins in the ab-plane in the Sr$_2$IrO$_4$ structure vanishing intensity is expected at the L$_2$ regardless of whether the insulating state emerges from a S=1/2 or ${J}_{\mathrm{eff}}\mathbf{=}1/2$ ground state \cite{PhysRevLett.112.026403}. While this adds further debate as to whether the ${J}_{\mathrm{eff}}\mathbf{=}1/2$ state exists in Sr$_2$IrO$_4$, we have shown that here the spins are, a least in certain Ru concentrations and temperature regions, aligned along the c-axis. In this case the magnetic structure does not contribute to the suppression of intensity of the L-edges, instead any suppression of intensity at the L$_2$ edge can be considered to be due to an alteration towards a ${J}_{\mathrm{eff}}\mathbf{=}1/2$ ground state. 
\begin{figure}[tb]
   \centering        
                        \includegraphics[trim=0cm 0.7cm 0.0cm 0.5cm,clip=true, width=0.8\columnwidth]{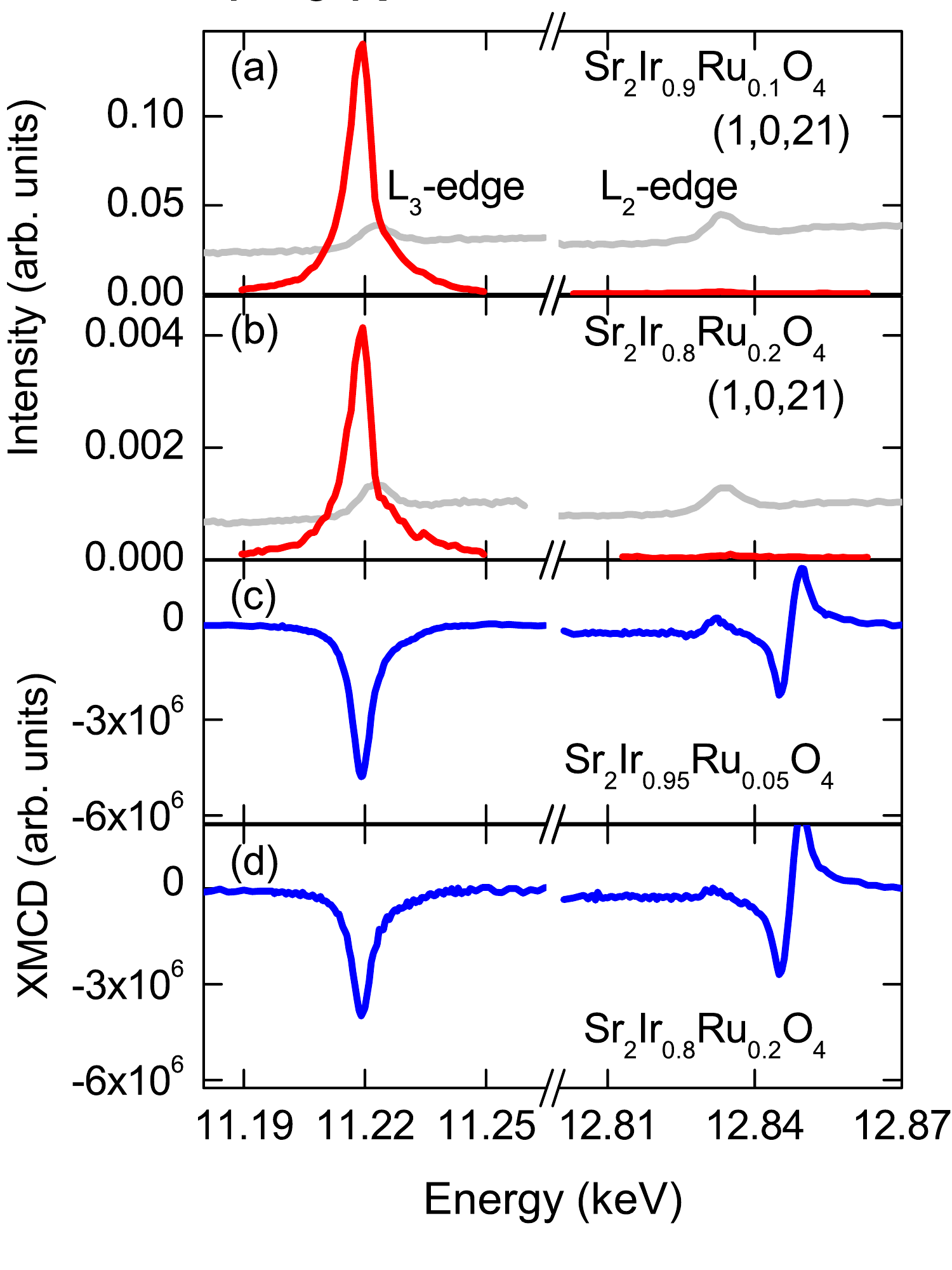}                    
                   \caption{\label{Escans_Ru0p1_0p2} Energy scans through the L$_2$ and L$_3$ edges using RMXS in (a) Sr$_2$Ir$_{0.9}$Ru$_{0.1}$O$_4$  and (b) Sr$_2$Ir$_{0.8}$Ru$_{0.2}$O$_4$ single crystals within the magnetically ordered phase (T= 5 K). The red and grey solid lines correspond to diffraction and absorption, respectively. XMCD energy scans in a $\pm$3 T field at 1.8 K for (c) Sr$_2$Ir$_{0.95}$Ru$_{0.05}$O$_4$ and (d) Sr$_2$Ir$_{0.8}$Ru$_{0.2}$O$_4$. The variation above 12.84 keV is an artifact of the field rather than from any XMCD signal from the sample.}
\end{figure}

We therefore focus on the (1,0,$2n$+1) reflections in Sr$_2$Ir$\rm _{1-x}$Ru$\rm _{x}$O$_4$ for x=0.1 and 0.2. The results at the (1 0 21) magnetic Bragg reflection at 5 K are shown in Fig.~\ref{Escans_Ru0p1_0p2} for both concentrations. At the L$_3$ edge we observe a large enhancement for both Sr$_2$Ir$_{0.9}$Ru$_{0.1}$O$_4$ and Sr$_2$Ir$_{0.8}$Ru$_{0.2}$O$_4$ in the $\sigma$-$\pi$ measurements, as expected for magnetic scattering. The maximum, as required, occurs at the inflection point of the absorption edge at 11.915 keV. Contrastingly the behavior at the L$_2$ edge shows very weak intensity in the  $\sigma$-$\pi$  measurements, with only weak scattering positioned at the absorption edge. This result therefore indicates that the SOC driven  ${J}_{\mathrm{eff}}\mathbf{=}1/2$ Mott insulating state exists within all of the magnetically order regimes of Sr$_2$Ir$\rm _{1-x}$Ru$\rm _{x}$O$_4$. This indicates that in general the ${J}_{\mathrm{eff}}\mathbf{=}1/2$ state can host a variety of structures and interactions.

To provide further evidence that the L-edge branching ratio is indeed a valid measurement of the existence of a ${J}_{\mathrm{eff}}\mathbf{=}1/2$  state we performed XMCD measurements on powder samples of x=0.05 (both ab-plane and c-axis ordering) and x=0.2 (only c-axis aligned spins) concentrations, see Fig.~\ref{Escans_Ru0p1_0p2}(c)-(d). These measurements do not rely on measuring at a magnetic Bragg reflection so in the x=0.05 it will probe a mixture of magnetic ordering whereas in x=0.2 it will probe only the c-axis ordering. The results show a reduced measured intensity with decreasing Ir concentrations, as would be expected. The branching ratio behavior is the same with the results showing a strongly suppressed signal at the $L_2$ edge in both samples indicating a strong role of SOC on the ground states.

\subsection{\label{sec:crystalstruct} Structural dependence of  Sr$_2$Ir$_{\rm 1-x}$Ru$\rm _x$O$_4$}

\begin{figure}[tb]
   \centering      
            \includegraphics[trim=1.2cm 6.6cm  2.5cm 0.2cm,clip=true, width=0.75\columnwidth]{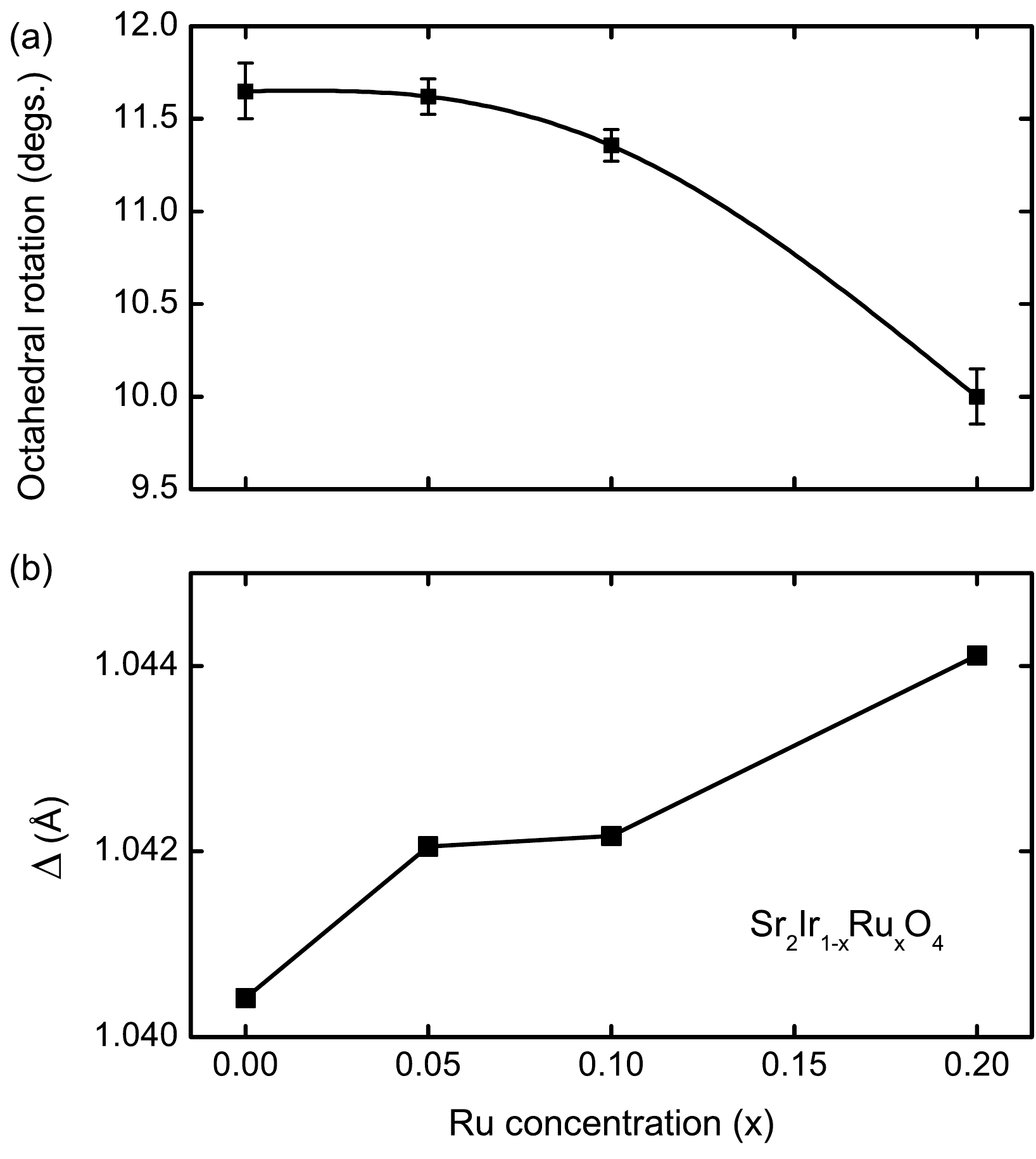}                    
                   \caption{\label{Str_Rudoped} Single crystal neutron diffraction on Sr$_2$Ir$\rm _{1-x}$Ru$\rm _{x}$O$_4$ for x=0, 0.05, 0.1 and 0.2 at 50 K. (a) The octahedral rotation in the ab-plane ($\alpha$). (b) Tetragonal distortion of the octahedra due to the difference between Ir-O bonds in the ab-plane and along the c-axis ($\Delta_{\rm Ir-O}$).}
                   \end{figure}

In the magnetically ordered region of the phase diagram (x$<$0.3) there is no structural symmetry change. However by performing single crystal neutron diffraction in the magnetically ordered regime we are able to follow the octahedral alterations, that are believed to be intimately entwined with the nature of the magnetic insulating state \cite{PhysRevLett.102.017205}.  The change in the rotation in the $ab$-plane, $\alpha$, and the tetragonal distortion, $\Delta_{\rm Ir-O}$, here defined as the c/a ratio of the two octahedral Ir-O bonds, are shown in Fig.~\ref{Str_Rudoped}. As expected the octahedral rotation angle decreases along the series as we approach the I4/mmm phase where $\alpha=0$. In the ${J}_{\mathrm{eff}}\mathbf{=}1/2$ limit it has been argued that the magnetic spins directly follow the canting within the ab-plane. This has been verified in Sr$_2$IrO$_4$ \cite{0953-8984-25-42-422202}. From neutron scattering the occurrence of the (0,0,$odd$) reflection indicates a canting of the spin in the ab-plane. The presence of the (0,0,5) reflection in 10$\%$ Ru doping shows that this persists upon doping and the orientation of the spin is consistent with following the rotation of the octahedra. As well as the octahedral rotation change the tetragonal distortion due to an elongation along the c-axis of the octahedra increases. The significance in this has been shown theoretically in Refs.~\onlinecite{PhysRevLett.102.017205}  and \onlinecite{FranchiniOctahedra} to be a route to a spin flop from a magnetic structure with spins in the ab-plane to a c-axis aligned AFM structure without an alteration of the ${J}_{\mathrm{eff}}\mathbf{=}1/2$ state. The evolution of the octahedra are at least qualitatively in line with this behavior. However, the value of c/a predicted in Ref.  \onlinecite{FranchiniOctahedra} was 1.09 and therefore appreciably higher than the $\Delta_{\rm Ir-O}$ value for even 20$\%$ doping. Therefore a structural route cannot be the sole reason for the observed spin flop. 

\subsection{\label{sec:PhaseDiag} Phase diagram of Sr$_2$Ir$_{\rm 1-x}$Ru$\rm _x$O$_4$}

\begin{figure}[tb]
   \centering                   
   \includegraphics[trim=2.1cm 3cm 3cm 3cm,clip=true, width=1.0\columnwidth]{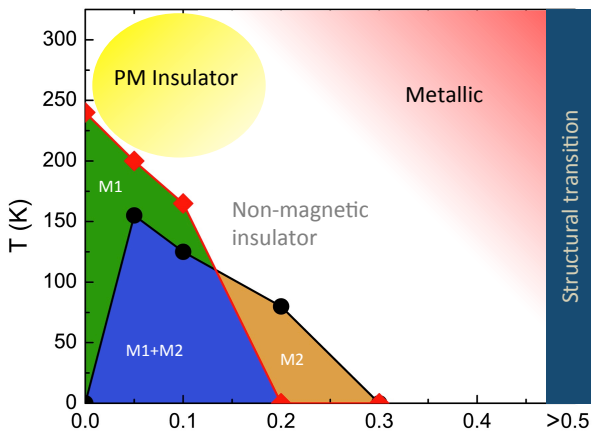} 
                   \caption{\label{FigPhaseDiag} Phase diagram for Sr$_2$Ir$\rm _{1-x}$Ru$\rm _{x}$O$_4$. The data points correspond to transition temperatures from single crystal neutron diffraction and RMXS. M1 denotes ab-plane magnetic ordering and M2 indicates c-axis ordering. The insulating regions are based on results presented in Ref. \onlinecite{PhysRevB.49.11890}.}
\end{figure}

Combining our neutron, x-ray and magnetization results allows us to construct a phase diagram for the series Sr$_2$Ir$\rm _{1-x}$Ru$\rm _{x}$O$_4$ before the structural phase transition at x$>$0.5, see Fig \ref{FigPhaseDiag}. Starting from the undoped x=0 insulator that undergoes magnetic order at 240 K the substitution of Ir$^{4+}$ for Ru$^{4+}$ leads to both a suppression of the MIT and an evolution of the magnetic structure. The x=0 magnetic structure (M1) is maintained up to x=0.1, a larger value than previous dopings with Mn or Rh  \cite{PhysRevB.86.220403,PhysRevB.89.054409}. However, a coexistence at low temperature between the x=0 basal plane ordering (M1) and the c-axis aligned magnetic structure (M2) exists for x=0.05 and x=0.1. Finally for x=0.2 only the M2 ordering is present, before the final removal of long range magnetic order at x$<$0.3.

\section{\label{sec:Discussion}Discussion}

The magnetic behavior of the series  Sr$_2$Ir$\rm _{1-x}$Ru$\rm _{x}$O$_4$ shows distinct behavior compared to the limited previous studies of doped Sr$_2$IrO$_4$, with an evolution and coexistence of long ranged magnetically ordered structures. This indicates the sensitivity and potential tuneabilty  of Sr$_2$IrO$_4$ to a variety of perturbations. For Mn-doped Sr$_2$IrO$_4$ the magnetic structure is the same as the M2 phase discussed here \cite{PhysRevB.86.220403}. For Rh-doped Sr$_2$IrO$_4$ the magnetic structure consists of spins in the ab-plane \cite{PhysRevB.89.054409}, as found for Sr$_2$IrO$_4$ in a small applied field \cite{KimScience}. For the case of  Rh-doped Sr$_2$IrO$_4$ there is no spin flop and  there exists a small region of short range correlations before the long range basal plane magnetic ordering sets in. We find no such regions in our investigations of Sr$_2$Ir$\rm _{1-x}$Ru$\rm _{x}$O$_4$.

To explain the coexistence of magnetic structures in Sr$_2$Ir$\rm _{1-x}$Ru$\rm _{x}$O$_4$ we considered and ruled out chemical phase separation, both in terms of structural phase separation and mixed valence. Moreover while the similarities to the ortho-G-AF phase from Kugel-Khomskii orbital ordering in Ref.~\onlinecite{PhysRevB.87.064407} is intriguing it remains unclear as to the validity of this model in Sr$_2$Ir$\rm _{1-x}$Ru$\rm _{x}$O$_4$.  Instead we find electronic phase separation to be the most consistent scenario for the coexistence of M1 and M2 based on our results and in analogy to separate investigations \cite{PhysRevB.89.054409, PhysRevB.89.104406, DhittalSr327Ru}. One potential cause of this phase separation, that manifests in distinct magnetic structures, is a competing and delicate balance of magnetic interactions in the Sr$_2$Ir$\rm _{1-x}$Ru$\rm _{x}$O$_4$ system due to altered exchange pathways. For example some bonds contain Ir-O-Ru-O-Ir bonds and others  will contain -Ir-O-Ir-O-Ir bonds that have different exchange interactions, extended to three dimensions. Since the correlation lengths of the two phases are not distinct then the phase separation is not limited to isolated small regions in the vicinity of the Ru dopant, but extends throughout the lattice. While the specific microscopic route to the spin flop transition is puzzling the alteration of the octahedra along with an introduction of anisotropy due to the dopant likely play a role. Despite the evolution and coexistence of magnetic structures the mechanism of the insulating state appears to remain unchanged and driven by the Mott mechanism splitting of the SOC enhanced ${J}_{\mathrm{eff}}\mathbf{=}1/2$ ground state.

Our results show that the Ir moments remain long ranged ordered up until Ru=0.3. The alternative doping of Rh in the series Sr$_2$Ir$_{1-x}$Rh$_x$O$_4$ instead found a removal of magnetic order at around half the concentration of Rh=0.17 \cite{PhysRevB.89.054409}. They considered a percolation driven suppression of magnetic order, in analogy to the cuprates, to most adequately describe the behavior. The large discrepancy between the concentration that magnetic order is suppressed between Rh and Ru doping at first glance appears to imply a divergence of behavior. However since the Rh introduced into Sr$_2$IrO$_4$ adopted the Rh$^{3+}$ valance it did not simply replace the Ir$^{4+}$ ions, but created two non-magnetic dopants (Rh$^{3+}$ and Ir$^{5+}$). Therefore that created an effective percolation value of 2x=0.34, that is close to the value we find for Sr$_2$Ir$\rm _{1-x}$Ru$\rm _{x}$O$_4$. Standard percolation theory predicts a value of x=0.4 as the concentration for the removal of magnetic order. So while the behavior for both Ru and Rh doped Sr$_2$IrO$_4$ is consistent with a percolation scenario, it appears to fall short of a full description. A full understanding incorporating further interactions such as SOC, 4d-5d magnetic interactions and band hybridization appears necessary to reproduce the observed behavior, with Ru doping potentially more favorable due to the direct replacement of Ir$^{4+}$ with Ru$^{4+}$ ions.

\section{\label{sec:Conclusion}Conclusion}

We have investigated the series Sr$_2$Ir$\rm _{1-x}$Ru$\rm _{x}$O$_4$ using both neutrons and resonant x-ray scattering to find the magnetic structure in the ordered regime of the phase diagram and assign the nature of the insulating state. Our results indicate a coexistence of  two doping induced magnetic structures up to 10$\%$ Ru substitution, compatible with an electronic phase separated system. At higher Ru concentration of 20$\%$ the magnetic structure consists solely of c-axis aligned spins, indicating a spin flop transition from the undoped Sr$_2$IrO$_4$ basal plane magnetic structure. Substituting in additional Ru removes long range magnetic order. We are able to use the RMXS L-edge branching ratio to assign the insulating behavior within the full magnetically ordered region of the phase diagram as consisting of a ${J}_{\mathrm{eff}}\mathbf{=}1/2$ SOC enhanced insulating state.

\begin{acknowledgments}
S.C. thanks A.~M.~Oles for illuminating discussions. This research at ORNL's High Flux Isotope Reactor was sponsored by the Scientific User Facilities Division, Office of Basic Energy Sciences, U.S. Department of Energy. Part of the work (AFM, CC, DM, BCS, GC) was supported by the U.S. Department of Energy, Office of Science, Basic Energy Sciences, Materials Sciences and Engineering Division. Use of the Advanced Photon Source, an Office of Science User Facility operated for the U.S. DOE Office of Science by Argonne National Laboratory, was supported by the U.S. DOE under Contract No. DE-AC02-06CH11357.
\end{acknowledgments}


%

\end{document}